\documentclass[draftcls,onecolumn]{IEEEtran}

\usepackage{amsmath}
\usepackage{amssymb}
\usepackage{amsfonts}
\usepackage{mathrsfs}
\usepackage{epsfig}
\usepackage{bm}
\usepackage{color}
\usepackage{cite}

\begin{document}

\newcommand{\tc}[2]{\textcolor[rgb]{#1}{#2}}
\newcommand{\ped}[1]{_{\text{#1}}}
\newcommand{\api}[1]{^{\text{#1}}}
\newcommand{\diff}[1]{\text{d}#1}
\newcommand{\pped}[1]{_{\scriptscriptstyle #1}}
\newcommand{\aapi}[1]{^{\scriptscriptstyle #1}}

\title{Direct Acceleration of Ions With Variable-frequency Lasers}

\author{F. Peano, J. Vieira, R. A. Fonseca, R. Mulas, G. Coppa, L. O. Silva
\thanks{F. Peano, J. Vieira, R. A. Fonseca, and L. O. Silva are with GoLP/Instituto de Plasmas e Fus\~ao Nuclear, Instituto 
Superior T\'ecnico, 1049-001 Lisboa, Portugal 
(e-mail: fabio.peano@ist.utl.pt, luis.silva@ist.utl.pt)}
\thanks{R. A. Fonseca is also with Departamento de Ci\^encias e Tecnologias da Informa\c{c}\~ao, Instituto Superior de Ci\^encias do Trabalho e da Empresa, 1649-026 Lisboa, Portugal}
\thanks{R. Mulas and G. Coppa are with Dipartimento di Energetica, Politecnico di Torino, 10129 Torino, Italy}
}

\maketitle

\begin{abstract}
A method is proposed for producing monoergetic, high-quality ion beams in vacuum, via direct acceleration by the electromagnetic field of two counterpropagating, variable-frequency lasers: ions are trapped and accelerated by a beat-wave structure with variable phase velocity, allowing for fine control over the energy and the charge of the beam via tuning of the frequency variation. The physical mechanism is described with a one-dimensional theory, providing the general conditions for trapping and scaling laws for the relevant features of the ion beam. Two-dimensional, electromagnetic particle-in-cell simulations, in which hydrogen gas is considered as an ion source, confirm the validity and the robustness of the method.
\end{abstract}

%\begin{IEEEkeywords}
%Laser-based acceleration, ion acceleration, direct acceleration, vacuum acceleration, optical trapping, optical acceleration
%\end{IEEEkeywords}

\section{Introduction}
\label{sec:intro}

The recent developments in the technology of ultraintense infrared (IR) lasers have opened the way to compact particle accelerators for ions and electrons. In the case of electrons, the possibility of achieving relativistic radiation intensities ($I\gtrsim10^{18}$ W/cm$^2$) motivated a number of proposals for direct acceleration in vacuum by the electromagnetic (EM) field of one or more laser pulses \cite{Esarey_1,vacuum_0,vacuum_1,vacuum_2,vacuum_3,vacuum_4,vacuum_5,vacuum_5b,vacuum_5c,vacuum_6,vacuum_7,vacuum_7b,vacuum_8}. However, indirect processes, in which the acceleration is obtained via laser-driven plasma waves \cite{wakefield_1}, were shown to be more effective and reliable \cite{wakefield_2,wakefield_3,wakefield_4,wakefield_5}. In the case of ions, for which relativistic radiation intensities ($I\gtrsim10^{24}$ W/cm$^2$) are still beyond the limits of current technology, direct acceleration has been mostly unexplored (only very recently, direct acceleration by tightly focused, radially polarized lasers has been investigated \cite{Salamin_OL}).
Most of the proposals for laser-based ion acceleration rely on indirect processes, in which the ions are accelerated by the space-charge field in laser-irradiated solid targets (cf. \cite{ion_accel_0,ion_accel_1,ion_accel_2,ion_accel_3,ion_accel_4,ion_accel_5,ion_accel_6,ion_accel_7,ion_accel_8,ion_accel_9} and, for a review, \cite{Borghesi,Mendonca,Bulanov}), resorting to various physical mechanisms, such as, with increasing laser intensity, plasma-expansions \cite{expansion_0,expansion_1,expansion_2}, electrostatic shocks \cite{Silva,Denavit}, and the laser-piston regime \cite{Esirkepov}.
In such processes, the properties of the accelerated ions are often difficult to control, making the production of high-quality beams a challenging task. 
In the spirit of conventional accelerators, better results would be obtained using schemes in which the acceleration is provided by a controllable wave structure. To accelerate ions in the nonrelativistic regime, a slow wave is required, whose velocity increases as the ions accelerate. Possible methods have been proposed to drive such waves in plasmas \cite{Katsouleas,Shvets_Wurtele_1,Shvets_Wurtele_2}.
Standing or slow waves can also be generated in the beating of counterpropagating lasers with equal or slightly different frequencies: a typical example is the optical injection of electrons into a plasma wave in a laser-wakefield accelerator (LWFA) \cite{wakefield_5,injection_2,injection_3}. The phase velocity of a beat wave can be modified in a natural way, by varying the frequency of the beating lasers (the use of two counterpropagating laser beams with a chirp has already been considered for the electron injection in a LWFA \cite{Esarey_private}, but relying on a different mechanism than the one explored here).

This Paper shows that, even at nonrelativistic radiation intensities ($I \ll 10^{24}$ W/cm$^2$, in the case of IR lasers), ions can be accelerated to high energies directly by the ponderomotive force of an EM beat wave \cite{injection_2,injection_3} with variable phase velocity, driven by two counterpropagating lasers with variable frequency \cite{Peano_NJP}. This  scheme for acceleration in vacuum does not rely on complex laser-matter couplings and works in a single-particle regime, as in conventional accelerators, offering the possibility of direct, efficient control over the relevant beam features (e.g., mean energy, energy spread, and total accelerated charge) via regulation of the fundamental parameters of the two lasers (i.e., intensity, duration, and frequency variation). The technique can operate in a wide frequency range, including the IR and visible region of the EM spectrum, where extreme intensities are available, and it can be adapted to any source of ions: it can serve as an energy booster, accelerating ions from a beam with a given energy, or it can accelerate ions starting at rest from a tenuous gas or vapor that is turned into a plasma by the laser radiation.

The Paper is organized as follows: in Section \ref{sec:theory}, the acceleration scheme is explored in detail with a one-dimensional theoretical model; in Section \ref{sec:scalings}, the scaling laws for the relevant ion beam properties are presented and the requirements in terms of the laser features are discussed; in Section \ref{sec:simulations}, numerical results from self-consistent simulations in two dimensions are presented. Finally, in Section \ref{sec:conclusions}, the conclusions are stated.

\section{Theory of direct ion acceleration}
\label{sec:theory}
\begin{figure}[!htb]
\centering \epsfig{file=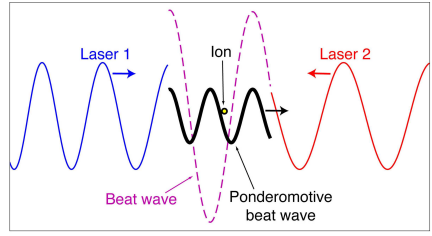, width=2.2in}
\caption{(Color online) Sketch of the acceleration technique: a test ion is trapped and accelerated by the slow ponderomotive beat wave generated by two counterpropagating, variable-frequency lasers. Thin lines represent the vector potential of the two lasers (solid lines) and the sum of the vector potentials in the superposition region (dashed line); the thick, solid line represents the ponderomotive beat-wave potential.}
\label{fig:scheme}
\end{figure}

The acceleration technique proposed here is analyzed in the framework of a relativistic, one-dimensional, single-particle theory, in which the amplitude of both lasers is constant, referring to the configuration sketched in Fig. \eqref{fig:scheme}. When considering the typical conditions required for the method to be effective, this model provides an accurate description of the fundamental features of the acceleration mechanism, as also confirmed by two-dimensional (2D) particle-in-cell (PIC) simulations (cf. Section \ref{sec:simulations}).

\subsection{Field configuration and basic equations}
\label{sec:equations}
In the present theory, the lasers are described as two coherent EM waves propagating along the $x$ direction, with vector potentials
\begin{equation}
{\bf A}_j = A_j\Big\{\cos\left(\theta\right)\sin\left[\Phi_j\left(\xi_j\right)\right]\hat{{\bf e}}_y + \sin\left(\theta\right)\cos\left[\Phi_j\left(\xi_j\right)\right]\hat{{\bf e}}_z\Big\} \text{,}
\label{eq:vect_pot}
\end{equation}   
where $\hat{{\bf e}}_y$ and $\hat{{\bf e}}_z$ are unit vectors, $\theta$ determines the polarization type (e.g., $\theta = \pi/2$ for linear polarization in the $z$ direction, $\theta = \pi/4$ for circular polarization), $j=1,2$ labels the lasers, $\xi_1 = x - ct$ (laser 1 propagates from left to right), $\xi_2 = -x-ct$ (laser 2 propagates from right to left), $A_j$ is the amplitude, and $\Phi_j\left(\xi_j\right)$ are two arbitrary functions with first derivative $\Phi_j^\prime\left(\xi_j\right)>0$ (henceforth, for the purpose of readability, the arguments $\xi_j$ will be omitted unless necessary). The local wavenumbers, $k_j\left(x,t\right)=\frac{\partial\Phi_j}{\partial x}$, and frequencies, $\omega_j\left(x,t\right)=-\frac{\partial\Phi_j}{\partial t}$, are $k_1\left(x,t\right)=\Phi_1^\prime$, $\omega_1\left(x,t\right)=c\Phi_1^\prime$, $k_2\left(x,t\right)=-\Phi_2^\prime$, and $\omega_2\left(x,t\right)=c\Phi_2^\prime$.
The superposition of the two EM waves can be described in terms of the two corresponding beat waves: (i) a fast, superluminal beat wave, having wavenumber $K(x,t)=\frac{1}{2}(\Phi_1^\prime-\Phi_2^\prime)$, frequency $\Omega(x,t)=\frac{c}{2}(\Phi_1^\prime+\Phi_2^\prime)$, and phase velocity $V_\phi(x,t)=\frac{\Omega}{K}=c\frac{\Phi_1^\prime+\Phi_2^\prime}{\Phi_1^\prime-\Phi_2^\prime}$, and (ii) a slow, subluminal beat wave, having wavenumber $k(x,t)=\frac{1}{2}(\Phi_1^\prime+\Phi_2^\prime)$, frequency $\omega(x,t)=\frac{c}{2}(\Phi_1^\prime-\Phi_2^\prime)$ and phase velocity $v_\phi(x,t)=\frac{\omega}{k}=c\frac{\Phi_1^\prime-\Phi_2^\prime}{\Phi_1^\prime+\Phi_2^\prime}$ (the following relations hold: $kc=\Omega$, $Kc=\omega$, and $v_\phi V_\phi = c^2$).
The ratio between the frequencies of the two laser beams in $x=0$ at $t=0$ is assumed to be such that the phase velocity of the slow beat wave matches the initial velocity of the ion, $c\beta_0$, i.e., such that $v_\phi(0,0)=c\beta_0$, leading to $\frac{\omega_2(0,0)}{\omega_1(0,0)}=\frac{1-\beta_0}{1+\beta_0}$.

By resorting to the conservation of the transverse canonical momentum, $p_y = -\frac{q}{c}\left({\bf A}_1+{\bf A}_2\right)\cdot\hat{{\bf e}}_y$ and $p_z = -\frac{q}{c}\left({\bf A}_1+{\bf A}_2\right)\cdot\hat{{\bf e}}_z$, the $x$ component of the equation of motion for the ion is written, in cgs units, as 
\begin{equation}
\frac{\diff{p_x}}{\diff{t}} = -\frac{q^2}{2\gamma Mc^2}\frac{\partial}{\partial x}\left({\bf A}_1+{\bf A}_2\right)^2 \text{,}
\label{eq:x_motion}
\end{equation} 
where $q$ and $M$ are the ion charge and mass, respectively, and $\gamma=\sqrt{1+\frac{p_x^2}{M^2c^2}+\frac{q^2\left({\bf A}_1+{\bf A}_2\right)^2}{M^2c^4}}$ is the Lorentz factor. 
As shown in the Appendix, if the radiation intensity is not so high as to drive transverse oscillations with relativistic velocities (i.e., if $\hat{A}_j = \frac{qA_j}{Mc^2}\ll1$), and as long as the characteristic time scales of the two beat waves can be separated, Eq. \eqref{eq:x_motion} can be averaged over the fast time scale to yield the ponderomotive equation of motion
\begin{align}
\dfrac{\diff{\hat{p}_x}}{\diff{\hat{t}}} & = -\dfrac{\hat{A}_1\hat{A}_2}{2\gamma}\dfrac{\partial}{\partial \hat{x}}\cos\left(\Phi_1-\Phi_2\right) \nonumber \\
& = \dfrac{\hat{A}_1\hat{A}_2}{\gamma}\hat{k}\left(\hat{x},\hat{t}\right)\sin\left(\Phi_1-\Phi_2\right)
\text{,}
\label{eq:x_motion_av}
\end{align}
where dimensionless quantities (denoted with hatted symbols henceforth) have been adopted: $\hat{k}=\frac{k}{k_0}$, with $k_0=k\left(0,0\right)$, $\hat{t} = k_0ct$, $\hat{x} = k_0x$, and $\hat{p}_x = \frac{p_x}{Mc}$ [in these units, the initial frequencies of the lasers are $\hat{\omega}_1(0,0)=1+\beta_0$ and $\hat{\omega}_2(0,0)=1-\beta_0$]. In Eq. \eqref{eq:x_motion_av}, the averaged Lorentz factor is defined as $\gamma^2 = 1+\hat{p}_x^2+\hat{A}_1^2 /2+\hat{A}_2^2 /2+\hat{A}_1\hat{A}_2\cos\left(\Phi_1-\Phi_2\right)$. Equation \eqref{eq:x_motion_av} can be obtained from the Hamiltonian $\mathscr{H}\left(\hat{x},\hat{p}_x,\hat{t}\right)=\big[1+\hat{p}_x^2+\hat{A}_1^2 /2+\hat{A}_2^2 /2+\hat{A}_1\hat{A}_2\cos\left(\Phi_1-\Phi_2\right)\big]^{1/2}$, using $\frac{\diff{\hat{p}_x}}{\diff{\hat{t}}}=-\frac{\partial\mathscr{H}}{\partial \hat{x}}$ (the other Hamilton equation being $\frac{\diff{\hat{x}}}{\diff{\hat{t}}}=\frac{\partial\mathscr{H}}{\partial \hat{x}}=\frac{\hat{p}_x}{\gamma}$); accordingly, the variation of the ion energy, $\frac{\diff{\gamma}}{\diff{\hat{t}}}=\frac{\partial\mathscr{H}}{\partial \hat{t}}$, is given by
\begin{align}
\dfrac{\diff{\gamma}}{\diff{\hat{t}}} & = \dfrac{\hat{A}_1\hat{A}_2}{2\gamma}\dfrac{\partial}{\partial \hat{t}}\cos\left(\Phi_1-\Phi_2\right) \nonumber \\
& = \dfrac{\hat{A}_1\hat{A}_2}{\gamma}\hat{\omega}\left(\hat{x},\hat{t}\right)\sin\left(\Phi_1-\Phi_2\right)
\text{.}
\label{eq:energy}
\end{align}
Equation \eqref{eq:energy} shows that the presence of a frequency variation is a necessary condition to effectively transfer energy from the beating EM waves to the ion over long times (cf., the Lawson-Woodward theorem \cite{Esarey_1,LW_1,LW_2}). If the frequency of both lasers is constant (i.e., if $\Phi_1^{\prime}$ and $\Phi_2^{\prime}$ are constants), Eq. \eqref{eq:energy} reduces to
\begin{equation}
\dfrac{\diff{\gamma}}{\diff{\hat{t}}}=\dfrac{\hat{A}_1\hat{A}_2}{\gamma}\beta_0\sin\left[2\left(\hat{x} - \beta_0\hat{t}\right) + \phi_0\right]
\text{,}
\label{eq:energy_2}
\end{equation}
where $\phi_0 = \Phi_1\left(0\right)-\Phi_2\left(0\right)$ is the initial phase of the beat wave. 
In the reference frame where $\beta_0=0$, $\mathscr{H}$ is time-indepedent, $\hat{p}_x^2+\hat{A}_1\hat{A}_2\cos\left(2\hat{x}+\phi_0\right)$ is constantly equal to $\hat{A}_1\hat{A}_2\cos\left(\phi_0\right)$ (i.e., variations in $\hat{p}_x^2$ are compensated by variations in $\hat{p}_y^2+\hat{p}_z^2$), $\hat{p}_x$ oscillates between $\pm\hat{P}$, with $\hat{P}=\{\hat{A}_1\hat{A}_2[1+\cos(\phi_0)]\}^{1/2}$, and $\gamma$ remains constantly equal to $\Gamma=[1+\hat{A}_1^2 /2+\hat{A}_2^2 /2+\hat{A}_1\hat{A}_2\cos(\phi_0)]^{1/2}$; in any other frame, $\mathscr{H}$ is time-depedent, $\hat{p}_x$ oscillates between $(\beta_0\Gamma\pm\hat{P})/(1-\beta_0^2)$, and $\gamma$ oscillates between $(\Gamma\pm\beta_0\hat{P})/(1-\beta_0^2)$. Thus, $\gamma$ cannot grow over long times.
However, if at least one of the lasers has variable frequency, causing the phase velocity of the ponderomotive beat wave to vary, $\mathscr{H}$ is time-depedent in any reference frame, and continuous energy transfer from the beating EM waves to the particle is possible. For appropriate choices of $\Phi_1$ and $\Phi_2$, the beat wave can trap the ion and accelerate it to large energies.  

\subsection{Resonant solutions}
\label{sec:resonant}
A resonant solution of Eq. \eqref{eq:x_motion_av}, $\hat{X}\left(\hat{t}\right)$, is defined by the resonance condition
\begin{equation}
\Phi_1\left[\hat{X}\left(\hat{t}\right)-\hat{t}\right]-\Phi_2\left[-\hat{X}\left(\hat{t}\right)-\hat{t}\right]=\phi_0
\text{,}
\label{eq:res_cond}
\end{equation}
which imposes exact phase-locking between the ion and the ponderomotive beat wave. In order to find a particular resonant solution, one can choose the expression for $\Phi_1$ (or $\Phi_2$) arbitrarily and then seek the expression of $\Phi_2$ (or $\Phi_1$) for which Eq. \eqref{eq:x_motion_av} and condition \eqref{eq:res_cond} are simultaneously satisfied. By making use of \eqref{eq:res_cond}, Eqs. \eqref{eq:x_motion_av} and \eqref{eq:energy} can be written, as
\begin{equation}
\dfrac{\diff{\hat{p}\pped{\parallel}}}{\diff{\tau\pped{\parallel}}} =\frac{\hat{A}_1\hat{A}_2}{2\gamma_{0{\scriptscriptstyle\perp}}^2}\sin\left(\phi_0\right)\left(\Phi_1^\prime+\Phi_2^\prime\right)
\label{eq:x_motion_av_tau}
\end{equation}
and
\begin{equation}
\dfrac{\diff{\gamma\pped{\parallel}}}{\diff{\tau\pped{\parallel}}} = \frac{\hat{A}_1\hat{A}_2}{2\gamma_{0{\scriptscriptstyle\perp}}^2}\sin\left(\phi_0\right)\left(\Phi_1^\prime-\Phi_2^\prime\right)
\text{,}
\label{eq:energy_tau}
\end{equation}
where $\gamma_{0{\scriptscriptstyle\perp}}^2=1+\hat{A}_1^2/2+\hat{A}_2^2/2+\hat{A}_1\hat{A}_2\cos\left(\phi_0\right)$, $\gamma\pped{\parallel}=\gamma/\gamma_{0{\scriptscriptstyle\perp}}=\left(1-\beta_x^2\right)^{-1/2}$, $\tau\pped{\parallel}=\gamma_{0{\scriptscriptstyle\perp}}\tau$ (with $\tau$ being the proper time, such that $\frac{\diff{\hat{t}}}{\diff{\tau}}=\gamma$), and $\hat{p}\pped{\parallel}=\hat{p}_x/\gamma_{0{\scriptscriptstyle\perp}}=\sqrt{\gamma\pped{\parallel}^2-1}$.
Assuming, without loss of generality, that $\Phi_2$ is known, Eqs. \eqref{eq:x_motion_av_tau} and \eqref{eq:energy_tau} can be combined to obtain
\begin{equation}
\dfrac{\diff{^2\hat{\xi}_1}}{\diff{\tau\pped{\parallel}}^2} =\mu_0\Phi_2^\prime\left[-\hat{\xi}_1-2\hat{t}\left(\tau\pped{\parallel}\right)\right]
\text{,}
\label{eq:x_motion_xi}
\end{equation}
where $\mu_0=\frac{\hat{A}_1\hat{A}_2}{\gamma_{0{\scriptscriptstyle\perp}}^2}\sin\left(\phi_0\right)$ and $\hat{\xi}_1=\hat{X}-\hat{t}$. By integrating Eq. \eqref{eq:x_motion_xi} one can determine the resonant ion trajectory in the parametric form $\left\{\hat{X}\left(\tau\pped{\parallel}\right),\hat{t}\left(\tau\pped{\parallel}\right)\right\}$ and then use Eq. \eqref{eq:res_cond} to determine $\Phi_1$. For a generic shape of $\Phi_2$, this process usually requires a numerical approach, because of the intricate dependence of $\hat{t}$ on $\tau\pped{\parallel}$ through the integral $\hat{t}=\int_0^{\tau\pped{\parallel}}\gamma\pped{\parallel}\left(\tau^\prime\pped{\parallel}\right)\diff{\tau^\prime\pped{\parallel}}$. In the special case in which the frequency of laser 2 is constant (as in  \cite{Peano_NJP}), i.e., $\Phi_2^\prime=1-\beta_0$, the right-hand side of Eq. \eqref{eq:x_motion_xi} does not depend on $\hat{t}$ and the calculation can be performed analytically. Integration of Eq. \eqref{eq:x_motion_xi} yields $\hat\xi_1\left(\tau\pped{\parallel}\right) = \frac{c_0}{2}\left(\mu \tau\pped{\parallel}-2\right)\tau\pped{\parallel}$, where $\mu = \mu_0/\gamma_{0{\scriptscriptstyle\parallel}}$, with $\gamma_{0{\scriptscriptstyle\parallel}}=\sqrt{1-\beta_0^2}$, and $c_0 = \gamma_{0{\scriptscriptstyle\parallel}}\left(1-\beta_0\right) = \sqrt{\frac{1-\beta_0}{1+\beta_0}}$; hence, $\hat{p}\pped{\parallel}\left(\tau\pped{\parallel}\right)$ and $\gamma\pped{\parallel}\left(\tau\pped{\parallel}\right)$ can be expressed as
\begin{equation}
\hat{p}\pped{\parallel}\left(\tau\pped{\parallel}\right)=\frac{\diff{\hat{X}}}{\diff{\tau\pped{\parallel}}}=\frac{1-c_0^2\left(1-\mu\tau\pped{\parallel}\right)^2}{2c_0\left(1-\mu\tau\pped{\parallel}\right)}
\text{,}
\label{eq:p_par}
\end{equation}
\begin{equation}
\gamma\pped{\parallel}\left(\tau\pped{\parallel}\right)=\frac{\diff{\hat{t}}}{\diff{\tau\pped{\parallel}}}=\frac{1+c_0^2\left(1-\mu\tau\pped{\parallel}\right)^2}{2c_0\left(1-\mu\tau\pped{\parallel}\right)}
\text{,}
\label{eq:gamma_par}
\end{equation}
and, consequently, $\hat{X}\left(\tau\pped{\parallel}\right)$ and $\hat{t}\left(\tau\pped{\parallel}\right)$ are
\begin{equation}
\hat{X}\left(\tau\pped{\parallel}\right)=\frac{1}{4\mu c_0}\left[c_0^2\left(\mu\tau\pped{\parallel}-2\right)\mu\tau\pped{\parallel}-2\log\left(1-\mu\tau\pped{\parallel}\right)\right]
\text{,}
\label{eq:X_res}
\end{equation}
\begin{equation}
\hat{t}\left(\tau\pped{\parallel}\right)=-\frac{1}{4\mu c_0}\left[c_0^2\left(\mu\tau\pped{\parallel}-2\right)\mu\tau\pped{\parallel}+2\log\left(1-\mu\tau\pped{\parallel}\right)\right]
\text{.}
\label{eq:t_res}
\end{equation}
Finally, by noticing that $1-\mu\tau\pped{\parallel}=\sqrt{1+\frac{2\mu}{c_0}\hat{\xi}_1}$, $\hat{X}$ and $\hat{t}$ can be written as functions of $\hat{\xi}_1$ and replaced in Eq. \eqref{eq:res_cond}, yielding
\begin{equation}
\Phi_1\left(\hat{\xi}_1\right)=\phi_0+\frac{1}{2\mu_0}\log\left[1+2\mu_0\left(1+\beta_0\right)\hat{\xi}_1\right]
\text{.}
\label{eq:res_cond_xi}
\end{equation}

It is useful to analyze the behavior of the resonant solution, given by Eqs. \eqref{eq:p_par}-\eqref{eq:t_res}, in the early stage of the acceleration, $\hat{t}\ll\mu^{-1}$, and in the asymptotic limit, $\hat{t}\gg\mu^{-1}$. For $\hat{t}\ll\mu^{-1}$, Eq. \eqref{eq:t_res} gives, to second order in $\hat{t}$, $\tau\pped{\parallel}\left(\hat{t}\right)\approx \frac{1}{\gamma_{0{\scriptscriptstyle\parallel}}}\hat{t}-\frac{\beta_0}{2\gamma_{0{\scriptscriptstyle\parallel}}^2}\mu\hat{t}^2$ and the explicit dependence of $\hat{p}\pped{\parallel}$, $\gamma\pped{\parallel}$, and $\hat{X}\pped{\parallel}$ on $\hat{t}$ is written as $\hat{p}\pped{\parallel}\left(\hat{t}\right) \approx \gamma_{0{\scriptscriptstyle\parallel}}\beta_0+\mu\hat{t}+\frac{1}{2\gamma_{0{\scriptscriptstyle\parallel}}}\mu^2\hat{t}^2$, $\gamma\pped{\parallel}\left(\hat{t}\right) \approx \gamma_{0{\scriptscriptstyle\parallel}}+\beta_0\mu\hat{t}+\frac{1+\gamma_{0{\scriptscriptstyle\parallel}}\beta_0}{2\gamma_{0{\scriptscriptstyle\parallel}}}\mu^2\hat{t}^2$, and $\hat{X}\left(\hat{t}\right) \approx \beta_0\hat{t} + \frac{1}{2\gamma_{0{\scriptscriptstyle\parallel}}^3}\mu\hat{t}^2$. In the nonrelativistic limit ($\beta_0 \rightarrow 0$, $\gamma_{0{\scriptscriptstyle\parallel}} \rightarrow 1$), the ion accelerates uniformly and its energy grows quadratically in time as $\frac{1}{2}\mu_0^2\hat{t}^2$. For $\hat{t}\gg\mu^{-1}$, the behavior of the solution depends on whether laser 1 and the ion are copropagating ($\mu>0$) or counterpropagating ($\mu<0$): if $\mu>0$, Eq. \eqref{eq:t_res} gives, to leading order in $\hat{t}$, $\tau\pped{\parallel}\left(\hat{t}\right) \approx \frac{1}{\mu} \left[1-\exp\left(\frac{c_0^2}{2}-2c_0\mu \hat{t}\right)\right]$ and, consequently, $\hat{X}\left(\hat{t}\right) \approx \hat{t} - \frac{c_0}{2\mu}$ and $\gamma\pped{\parallel}\left(\hat{t}\right) \approx \hat{p}\pped{\parallel}\left(\hat{t}\right) \approx \frac{1}{2c_0}\exp\left(2c_0\mu \hat{t}-\frac{c_0^2}{2}\right)$; if $\mu<0$, Eq. \eqref{eq:t_res} yields $\tau\pped{\parallel}\left(\hat{t}\right) \approx 2\sqrt{\frac{\hat{t}}{|\mu|c_0}}$ and, consequently, $\hat{X}\left(\hat{t}\right) \approx -\hat{t}$ and $\gamma\pped{\parallel}\left(\hat{t}\right) \approx \hat{p}\pped{\parallel}\left(\hat{t}\right) \approx \sqrt{c_0|\mu|\hat{t}}$. The behavior is asymmetric because, if $\mu>0$, the frequency must be increased to maintain phase-locking, causing a continuous increase of the ponderomotive force, whereas, if $\mu<0$, the frequency must be decreased, causing a continuous decrease of the ponderomotive force.

\subsection{Trapping}
\label{sec:trapping}
\begin{figure}[!htb]
\centering \epsfig{file=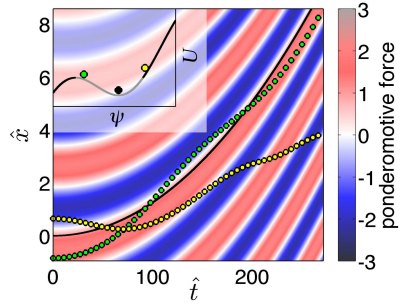, width=2.2in}
\caption{(Color online) Space-time evolution of the dimensionless ponderomotive force $-\frac{\partial}{\partial{\hat{x}}}\cos\left(\Phi_1-\Phi_2\right)$ (color scale, red and blue denoting accelerating and decelerating regions, respectively), and trajectories of a resonant ion (solid line), a non-resonant, trapped ion (green markers), and a non-resonant, untrapped ion (yellow markers), for a situation in which $\beta_0=0$, $\Phi_2^\prime = 1$, and $\Phi_1$ is given by expression \eqref{eq:res_cond_xi}, with $\hat{A}_1\hat{A}_2=5\times10^{-4}$ and $\phi_0=5\pi/6$. The inset represents the effective potential $U$ at $\hat{t}=0$, gray marking the trapping region, and the corresponding initial positions of the three ions.}
\label{fig:trajectories}
\end{figure}

Effective ion acceleration can occur also when $\Phi_1$ and $\Phi_2$ do not match the resonant solution (exact phase-locking). In fact, it is sufficient to choose $\Phi_1$ and $\Phi_2$ appropriately in order to guarantee that the ion trajectory stays close to the beat-wave trajectory until reaching the desired energy, i.e., that the ion is trapped by the beat wave.
For given $\Phi_1$ and $\Phi_2$, the beat-wave trajectory $\hat{x}_{\phi_0}\left(\hat{t}\right)$ (namely, the trajectory of the point of the beat wave having constant phase $\phi_0$) is determined by solving the equation $\Phi_1\left(\hat{x}_{\phi_0}-\hat{t}\right)-\Phi_2\left(-\hat{x}_{\phi_0}-\hat{t}\right)=\phi_0$ with respect to $\hat{x}_{\phi_0}$ [for resonant trajectories, $\hat{X}\left(\hat{t}\right)=\hat{x}_{\phi_0}\left(\hat{t}\right)$]. In order to determine the conditions for the occurrence of trapping, Eq. \eqref{eq:x_motion_av} is expressed in terms of the phase difference between particle and wave, $\psi=2\left[\hat{x}-\hat{x}_{\phi_0}\left(\hat{t}\right)\right]$, and using the proper time $\tau$, as
\begin{equation}
\frac{\diff^2{\psi}}{\diff{\tau^2}} = -\frac{\partial}{\partial\psi}U\left(\psi,\tau\right)
\text{,}
\label{eq:dpsi_dtau}
\end{equation} 
in which
\begin{equation}
U\left(\psi,\tau\right) =  2\hat{A}_1\hat{A}_2 \cos\left\{ \Phi_1\left[\hat{\xi}_{1,\phi_0}\left(\tau\right)+\frac{\psi}{2}\right]-\Phi_2\left[\hat{\xi}_{2,\phi_0}\left(\tau\right)-\frac{\psi}{2}\right]\right\} + 2\alpha_{\phi_0}\left(\tau\right)\psi
\text{,}
\label{eq:U}
\end{equation} 
where $\hat{\xi}_{1,\phi_0}$, $\hat{\xi}_{2,\phi_0}$, and $\alpha_{\phi_0}$ depend on $\tau$ through $\hat{t}$ as $\hat{\xi}_{1\phi_0}\left(\tau\right)=\hat{x}_{\phi_0}\left(\hat{t}\right)-\hat{t}$, $\hat{\xi}_{2,\phi_0}\left(\tau\right)=-\hat{x}_{\phi_0}\left(\hat{t}\right)-\hat{t}$, and $\alpha_{\phi_0}\left(\tau\right)=\frac{\diff^2}{\diff{\tau^2}}\hat{x}_{\phi_0}\left(\hat{t}\right)=\gamma\left(\hat{t}\right)\frac{\diff{}}{\diff{\hat{t}}}\left[\gamma\left(\hat{t}\right)\frac{\diff{}}{\diff{\hat{t}}}\hat{x}_{\phi_0}\left(\hat{t}\right)\right]$. According to Eq. \eqref{eq:dpsi_dtau}, trapping is allowed only if the frequency variation is slow enough to guarantee that the effective potential $U$ presents local minima.
For trapped ions, $|\psi|\ll 2\hat{\xi}_{j\phi_0}$ and $U$ can be approximated as
\begin{equation}
U\left(\psi,\tau\right) \approx 2\hat{A}_1\hat{A}_2\cos\left\{\hat{k}\left[\hat{x}_{\phi_0}\left(\hat{t}\right),\hat{t}\right]\psi +\phi_0\right\} + 2\alpha_{\phi_0}\left(\hat{t}\right)\psi
\label{eq:U2}
\end{equation}  
leading to the necessary condition for ion trapping
\begin{equation}
\left|\alpha_{\phi_0}\left(\hat{t}\right)\right|<\alpha\pped{\text{M}}\left(\hat{t}\right)=\hat{A}_1\hat{A}_2\hat{k}\left[\hat{x}_{\phi_0}\left(\hat{t}\right),\hat{t}\right]
\text{,}
\label{eq:trap_cond}
\end{equation}
where $\alpha\pped{\text{M}}\left(\hat{t}\right)$, is the maximum value of the ponderomotive force. Thus, trapping is allowed only if the maximum ponderomotive force of the beat wave is greater than the inertial force associated with the beat-wave acceleration.
The particular situation in which $\left|\alpha_{\phi_0}\left(\hat{t}\right)\right|=\alpha\pped{\text{M}}\left(\hat{t}\right)$ corresponds to a resonant solution in the particular case $\phi_0 = \pm \pi/2$. When $\Phi_1$ and $\Phi_2$ correspond to a resonant solution and $|\phi_0| \neq \pi/2$, regions where trapping is possible always exist, although the solution is stable for $\cos\left(\phi_0\right)<0$ and unstable for $\cos\left(\phi_0\right)>0$, with $\psi=0$ corresponding to the bottom and top of a potential well, respectively. 
The relative width of the potential well in $U\left(\psi,0\right)$, $\Delta\psi$, can be used to estimate the trapping efficiency $\eta\ped{tr}$ (defined as the fraction of trapped ions) as $\eta\ped{tr}\approx\frac{\Delta\psi}{2\pi}\simeq 1-\frac{2}{\pi}\arcsin[\alpha_{\phi_0}\left(0\right)/\alpha\pped{\text{M}}]$.
Examples of typical trapped and untrapped trajectories are shown in Fig. \ref{fig:trajectories}.

\subsection{Ion acceleration with linearly chirped lasers}
\label{eq:linear_chirp}
\begin{figure}[!htbp]
\centering \epsfig{file=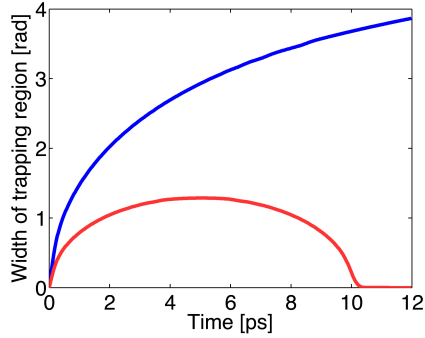, width=2.2in}
\caption{(Color online) Width of trapping region as a function of time when only laser one is chirped with $\sigma_1 = -A_1A_2$ (case A, blue/dark curve), and when both lasers are chirped with $\sigma_1 = -0.65A_1A_2$ and $\sigma_2=0.35A_1A_2$ (case B, red/light curve). Ion type: proton; $I_1=I_2=10^{20}\text{ W}/\text{cm}^2$.}
\label{fig:DPSI}
\end{figure}

\begin{figure}[!htbp]
\centering \epsfig{file=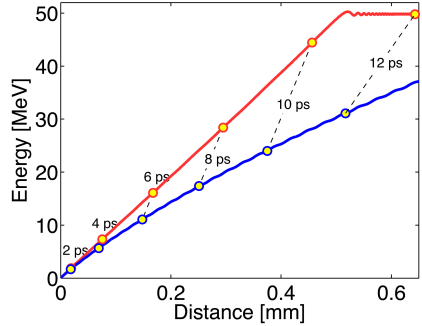, width=2.2in}
\caption{(Color online) Evolution of ion energy as a function of the acceleration distance for the same configurations, case A and case B, as in Fig. \ref{fig:DPSI}. Circles indicate equal time instants for both cases.}
\label{fig:x-en}
\end{figure}

An important case of variable-frequency lasers, particularly relevant in experiments, is that of linearly chirped beams, in which $\Phi_1$ and $\Phi_2$ take the form $\Phi_1\left(\hat{\xi}_1\right)=\phi_{01}+\left(1+\beta_0\right)\hat{\xi}_1+\sigma_1\hat{\xi}_1^2$ and $\Phi_2\left(\hat{\xi}_2\right)=\phi_{02}+\left(1-\beta_0\right)\hat{\xi}_2+\sigma_2\hat{\xi}_2^2$, where $\phi_{0j}$ are constant phases and $\sigma_j$ are the chirp coefficients. The local frequencies and wavenumbers are given by $\hat{\omega}_1\left(\hat{x},\hat{t}\right)=\hat{k}_1\left(\hat{x},\hat{t}\right)=1+\beta_0+2\sigma_1\hat{\xi}_1$ and $\hat{\omega}_2\left(\hat{x},\hat{t}\right)=-\hat{k}_2\left(\hat{x},\hat{t}\right)=1-\beta_0+2\sigma_2\hat{\xi}_2$. The wavenumber and the frequency of the slow beat wave are $\hat{k}\left(\hat{x},\hat{t}\right)=1+\sigma\pped{-}\hat{x}-\sigma\pped{+}\hat{t}$ and $\hat{\omega}\left(\hat{x},\hat{t}\right)=\beta_0+\sigma\pped{+}\hat{x}-\sigma\pped{-}\hat{t}$, where $\sigma\pped{-}=\sigma_1-\sigma_2$ and $\sigma\pped{+}=\sigma_1+\sigma_2$. The phase of the ponderomotive beat wave is $\Phi_1-\Phi_2=\phi_0+2\left(\hat{x}-\beta_0\hat{t}-\sigma\pped{+}\hat{x}\hat{t}\right)+\sigma\pped{-}\left(\hat{x}^2+\hat{t}^2\right)$, with $\phi_0 = \phi_{01}-\phi_{02}$. Consequently, the beat-wave trajectory can be written as $\hat{x}_{\phi_0}\left(\hat{t}\right)$ as
\begin{equation} 
\hat{x}_{\phi_0}\left(\hat{t}\right)=\frac{\sigma\pped{+}}{\sigma\pped{-}}\hat{t}-\frac{1}{\sigma\pped{-}}\left[1-\sqrt{1-2\left(\sigma\pped{+}-\beta_0\sigma\pped{-}\right)\hat{t}+\left(\sigma\pped{+}^2-\sigma\pped{-}^2\right)\hat{t}^2}\right]
\text{.}
\label{eq:x_phi0}
\end{equation} 
If only one laser is chirped, i.e., if $\sigma_2=0$ or $\sigma_1=0$, Eq. \eqref{eq:x_phi0} reduces, respectively, to
\begin{equation} 
\hat{x}^{(1)}_{\phi_0}\left(\hat{t}\right)=\hat{t}-\frac{1}{\sigma_1}\left[1-\sqrt{1-2\sigma_1\left(1-\beta_0\right)\hat{t}}\right]
\text{,}
\label{eq:x_phi0_1}
\end{equation}
\begin{equation} 
\hat{x}^{(2)}_{\phi_0}\left(\hat{t}\right)=-\hat{t}+\frac{1}{\sigma_2}\left[1-\sqrt{1-2\sigma_2\left(1+\beta_0\right)\hat{t}}\right]
\text{.}
\label{eq:x_phi0_2}
\end{equation}
Equations \eqref{eq:x_phi0}-\eqref{eq:x_phi0_2} suggest that the acceleration can be improved by chirping both lasers, with $\sigma_1\sigma_2<0$. When doing so, the specific values of $\sigma_1$ and $\sigma_2$ must be chosen with some care because the chirp in the two lasers affects the trapping efficiency in a different way.
This is readily seen by writing the inertial term $\alpha_{\phi_0}\left(\hat{t}\right)$ in the case of ions trajectories close to $\hat{x}^{(1)}_{\phi_0}\left(\hat{t}\right)$ and $\hat{x}^{(2)}_{\phi_0}\left(\hat{t}\right)$: for ions traveling with approximately the beat-wave velocity, $\alpha_{\phi_0}\left(\hat{t}\right) \approx \gamma_{\phi_0}\left(\hat{t}\right)\frac{\diff{}}{\diff{\hat{t}}}\left[\gamma_{\phi_0}\left(\hat{t}\right)\frac{\diff{}}{\diff{\hat{t}}}\hat{x}_{\phi_0}\left(\hat{t}\right)\right]$, where $\gamma_{\phi_0}$ is the Lorentz factor associated with the beat-wave trajectory $\hat{x}_{\phi_0}$, yielding
\begin{equation} 
\hat{\alpha}^{(1)}_{\phi_0}\left(\hat{t}\right)\approx\frac{-\sigma_1\sqrt{1-2\sigma_1\left(1-\beta_0\right)\hat{t}}}{\left[\beta_0-1+2\sqrt{1-2\sigma_1\left(1-\beta_0\right)\hat{t}}\right]^2}
\text{,}
\label{eq:x_aphi0_1}
\end{equation}
\begin{equation} 
\hat{\alpha}^{(2)}_{\phi_0}\left(\hat{t}\right)\approx\frac{\sigma_2\sqrt{1-2\sigma_2\left(1+\beta_0\right)\hat{t}}}{\left[1+\beta_0-2\sqrt{1-2\sigma_2\left(1+\beta_0\right)\hat{t}}\right]^2}
\text{.}
\label{eq:x_aphi0_2}
\end{equation}
According to Eqs. \eqref{eq:x_aphi0_1} and \eqref{eq:x_aphi0_2}, when the ion and the chirped laser are copropagating (e.g., $\sigma_1<0$ and $\sigma_2=0$), $\alpha_{\phi_0}\left(\hat{t}\right)$ is monotonically decreasing in time, causing the trapping regions to widen; on the contrary, when the ion and the chirped laser are counterpropagating (e.g., $\sigma_1=0$ and $\sigma_2>0$), $\alpha_{\phi_0}\left(\hat{t}\right)$ is monotonically increasing in time, causing the trapping regions to narrow.  
When chirping both lasers ($\sigma_1 \neq 0$ and $\sigma_2 \neq 0$), the use of a stronger chirp in the laser copropagating with the ion is in general convenient: in such a situation, the general expression for $\alpha_{\phi_0}\left(\hat{t}\right)$ has a minimum for a given time (which is $\hat{t}=0$ if $\sigma_1=-\sigma_2$), such that the width of the trapping regions increases until $\alpha_{\phi_0}\left(\hat{t}\right)$  reaches its minimum, and starts decreasing immediately after. This is illustrated in Fig. \ref{fig:DPSI}, showing the amplitude of the trapping region, $\Delta\psi$, as a function of time when only one laser is chirped (case A) and when both lasers are chirped (case B). In case A, $\Delta\psi$ grows monotonically towards its maximum value, $\Delta\psi=2\pi$. In case B, $\Delta\psi$ reaches a maximum and then starts decreasing; when $\Delta\psi$ vanishes the ion loses the beat wave and the acceleration stops. The corresponding evolution of the ion energy as a function of the acceleration distance is shown in Fig. \ref{fig:x-en}: in case A, the ion acceleration continues indefinitely, but with decreasing accelerating gradient; in case B, the ion accelerates with approximately constant accelerating gradient until losing the synchronization with the beat wave.

The maximum energy gain, $\Delta \epsilon\pped{\text{M}}$, occurs when the maximum chirp that allows for trapping is used, hence, according to Eq. \eqref{eq:x_phi0}, when $|\sigma\pped{-}|=\hat{A}_1\hat{A}_2$.
For non-relativistic ions, the maximum energy is then $\Delta \hat{\epsilon}\pped{\text{M}}\approx\frac{1}{2}\left(\frac{\diff{}}{\diff{\hat{t}}}\hat{x}_{\phi_0}\right)^2\approx\frac{1}{2}\left(\hat{A}_1\hat{A}_2\hat{t}\right)^2$, leading to the scaling law
\begin{align}
\Delta \epsilon\pped{\text{M}}\left[\text{MeV}\right] \approx & \ 0.8 \ Z\ped{i}^4A\ped{i}^{-3} I_1\left[10^{20} \ \mathrm{W/cm^2}\right]I_2\left[10^{20} \ \mathrm{W/cm^2}\right] \nonumber \\
& \  \times\lambda_{01}\left[\mu\text{m}\right]\lambda_{02}\left[\mu\text{m}\right] \Delta T^2\left[\text{ps}\right]\text{,}
\label{eq:EI_scaling}
\end{align}
where $A\ped{i}$ is the ion mass number and $\lambda_{0j}=2\pi/|k_j\left(0,0\right)|$ are the initial laser wavelengths. Similarly, the acceleration distance, $\Delta\hat{x}\approx \left|x_{\phi_0}\right|\approx\frac{1}{2}\hat{A}_1\hat{A}_2\hat{t}^2$, scales as
\begin{align}
\Delta x\left[\mu\text{m}\right] \approx & \ 6 Z\ped{i}^2A\ped{i}^{-2} I_1^{1/2}\left[10^{20} \ \mathrm{W/cm^2}\right]I_2^{1/2}\left[10^{20} \ \mathrm{W/cm^2}\right] \nonumber \\
& \  \times\lambda_{01}\left[\mu\text{m}\right]\lambda_{02}\left[\mu\text{m}\right] \Delta T^2\left[\text{ps}\right]\text{.}
\label{eq:x_scaling}
\end{align}  
By varying $\sigma_1$ and $\sigma_2$, the energy gain $\Delta \epsilon$ can be controlled, according to $\Delta \epsilon = \left(\frac{\sigma\pped{-}}{\hat{A}_1\hat{A}_2}\right)^2\Delta \epsilon\pped{\text{M}}$. When increasing $|\sigma\pped{-}|$, $\alpha_{\phi_0}$ increases and, consequently, the trapping efficiency $\eta\ped{tr}$ decreases; hence, the chirp coefficients must be tuned in order to obtain the desired trade-off between energy gain and total accelerated charge.

\section{Beam properties and laser requirements}
\label{sec:scalings}
By analyzing the range of validity of the present model, conclusions can be drawn for more general situations, in which spatial distributions of ions are considered and the effects due to the finite-size of the laser pulses are taken into account.

In the presence of a spatial distribution of ions, the scheme remains effective if the space-charge field due to the ion density $n\ped{i}$ does not perturb the EM beat-wave structure. Assuming that no neutralizing electron background is present, this establishes the criterion
\begin{equation}
n\ped{i}\left[10^{19} \ \text{cm}^{-3}\right] \ll Z\ped{i}^{-1}I_1^{1/2}\left[10^{20} \  \mathrm{W/cm^2}\right]I_2^{1/2}\left[10^{20} \ \mathrm{W/cm^2}\right]\text{.}
\label{eq:density}
\end{equation}
When taking into account realistic laser beams, having finite duration $T_j$ and spot size $W_{0j}$, both the longitudinal and the transverse dependence of the laser envelope must be considered.
As far as the longitudinal dependence is concerned, since long pulses (e.g., $T_j\gtrsim$ ps with visible/near-IR lasers) are required to obtain significant energy gains, the ponderomotive force associated with longitudinal intensity variations is negligible when compared  with the ponderomotive force of the beat wave. As a result, $\hat{A}_1$ and $\hat{A}_2$ can be considered as slowly varying functions of $\hat{t}$, and the only relevant effect is that intensities high enough for ion trapping exist only during a fraction of the overlap time for the two laser pulses; consequently, the duration of the acceleration process is limited to a fraction of the pulse duration, determined by the specific chirp law.
As an effect of the transverse variations of the radiation intensities, the ions are pushed outward by the transverse ponderomotive force of the laser pulses, whose magnitude scales as $A_j^2/W_{0j}$; hence, if the intensities of the two lasers are comparable, the ratio between the transverse forces and the ponderomotive force scales as $\frac{\lambda_0}{2\pi W_{0j}}\ll1$, with $\lambda_0=2\pi/k\left(0,0\right)$. In these conditions, the transverse cross section of the accelerating region is on the order of $\pi W_0^2$, where $W_0$ is the overlapped spot size of the lasers, and the length of the acceleration process is always limited to distances on the order of $z\ped{r}=\min\left(z\ped{r1},z\ped{r2}\right)$, where $z\ped{r$j$} = \pi W_{0j}^2/\lambda_{0j}$ are the Rayleigh lengths of the laser pulses;
therefore, the maximum total accelerated charge $Q\pped{\text{M}}$ (obtained when the laser duration is on the order of $z\ped{r}$) scales as
\begin{align}
Q\pped{\text{M}} \text{[pC]} \approx 5 \eta\ped{tr}Z\ped{i} n\ped{i}\left[10^{19}\text{ cm}^{-3}\right] W_0^2\left[\mu\text{m}\right]z\ped{r}\left[\mu\text{m}\right]\text{,}
\label{eq:Q_scaling}
\end{align}  
where $\eta\ped{tr}$ is the fraction of ions trapped by the beat wave (cf. Section \ref{sec:trapping}).

In light of these considerations, the scaling law \eqref{eq:EI_scaling} can be conveniently expressed in terms of the laser-pulse energies as
\begin{align}
\Delta \mathcal{E}\ped{M}[\mathrm{MeV}] \approx & \ 0.08 \frac{Z\ped{i}^4}{A\ped{i}^{3}} \frac{\mathcal{E}_1[\mathrm{J}]\mathcal{E}_2[\mathrm{J}]\lambda_{01}[\mu\mathrm{m}]\lambda_{02}[\mu\mathrm{m}]}{W_{01}^{2}[\mu\mathrm{m}]W_{02}^{2}[\mu\mathrm{m}]} \nonumber \\
\approx & \ 0.8 \frac{Z\ped{i}^4}{A\ped{i}^{3}} \frac{\mathcal{E}_1[\mathrm{J}]\mathcal{E}_2[\mathrm{J}]}{z\ped{r1}[\mu\mathrm{m}]z\ped{r2}[\mu\mathrm{m}]}\text{.}
\label{eq:EE_scaling}
\end{align}
Hence, if $W^2_{0j}/\lambda_{0j}$ (and hence $z\ped{r1}$) are fixed, the maximum energy gain depends only on the energy of the two laser pulses. For given pulse energies, wavelengths, and spot sizes, the combination of pulse lengths and intensities can be tuned in order to obtain the desired balance between the relative width of the final energy spectrum (which scales as $\frac{A_1A_2}{\Delta \mathcal{E}\ped{M}}$) and the total accelerated charge (which scales as $A_1A_2$ if the laser durations are longer than $z\ped{r}$, being constant otherwise). 
Furthermore, the scalings indicate that, for given pulse energies, and keeping $W_{0j}/\lambda_{0j}$ constant, higher energies are obtained when using lasers having shorter wavelengths. As an example, with 800 nm lasers (e.g., Ti:Sapphire lasers) focused at 10 $\mu$m, an energy gain $\Delta \mathcal{E}\ped{M}=1$ MeV (with protons) requires 440 J in each laser pulse; with 10 $\mu$m lasers (e.g., CO$_2$ lasers) and with the same $W_{0j}/\lambda_{0j}$, the same energy gain requires 5.5 kJ in each laser pulse.

Provided that the available laser energy is sufficient, the other fundamental requirement of the acceleration method is the frequency bandwidth. The total frequency excursions, $\Delta \omega_j$, necessary to accelerate an ion from the initial velocity $\beta_0$ to the final velocity $\beta$ are determined by 
\begin{equation}
\beta = \frac{2\beta_0+\Delta\omega_1-\Delta\omega_2}{2+\Delta\omega_1+\Delta\omega_2} \text{.}
\label{eq:domega}
\end{equation}
When only one laser is chirped, $\Delta \omega_j$ are given by $\Delta \omega_1=2\left(\beta-\beta_0\right)/\left(1-\beta\right)$ and $\Delta \omega_2=2\left(\beta-\beta_0\right)/\left(1+\beta\right)$. The minimum frequency excursion per laser is obtained when $\Delta \omega_1=-\Delta \omega_2=\beta-\beta_0$.  

The above estimates indicate that suitably chirped pulses, with energy level between 100 J and 1 kJ  and with frequency excursions on the order of 5-10\% of the central frequency (potentially available with current laser technology), should already allow for proof-of-principle experiments producing monoenergetic ion beams in the 100 keV - 1 MeV energy range. To achieve relativistic energies, highly energetic light sources capable of delivering coherent radiation over a large frequency range are required. Alternatives can be pursued, for instance, with multi-stage acceleration schemes.

\section{Numerical results}
\label{sec:simulations}
The effectiveness of the method and the validity of the estimates in Sec. \ref{sec:scalings} have been confirmed with one-dimensional (1D) and 2D, EM PIC simulations, using the Osiris 2.0 simulation framework \cite{osiris}, accounting self-consistently for space-charge and propagation effects.
Here, a selection of results is presented from a 2D simulation, in which the source of ions to be accelerated is provided simply by a slab of hydrogen, with a thickness of 75 $\mu$m and proton density $n\ped{i} = 5\times10^{16}$ cm$^{-3}$ (in the simulation, field ionization of the neutral hydrogen atoms is calculated by using the Ammosov-Delone-Krainov rates \cite{ADK}). 
Two approximately  gaussian laser pulses are employed, with spot sizes $W_{01}=W_{02}=10$ $\mu$m and durations $T_1=T_2=4.2$ ps, focused on the mid plane of the slab.
Laser 1, propagating from left to right, has peak intensity $I_1 = 1.3\times10^{21}$ W/cm$^2$ and it is linearly chirped according to $\Phi_1\left(\xi_1\right) = k_{01}\xi_1 + \sigma_1(\xi_1-\xi_{01})^2$, where $k_{01}=2\pi/\lambda_{01}$, with $\lambda_{01}=800$ nm, where $\xi_1$ indicates the distance from the pulse center, $\sigma_1=-2\times10^{-5} \ k_{01}^2$ is the chirp coefficient, and $\xi_{01}=6.25\times10^{3} \ k_{01}^{-1}$ is the point of the pulse where $k_1=k_{01}$, which needs to be specified when considering finite length pulses.
Laser 2, propagating from right to left, has peak intensity $I_2 = 8.5\times10^{20}$ W/cm$^2$, constant frequency, and wavelength $\lambda_{2}=800$ nm.
The two laser pulses are launched from opposite $x$ boundaries of the computational domain and hit the left face of the hydrogen slab at the same time. The computational domain is rectangular [$L_x=300$ $\mu$m ($5\times10^4$ cells) and $L_y=50$ $\mu$m ($100$ cells)], with open-space boundary conditions in $x$ and periodic boundary conditions in $y$. The left face of the hydrogen slab is located at $x=40$ $\mu$m.
\begin{figure}[!htbp]
\centering \epsfig{file=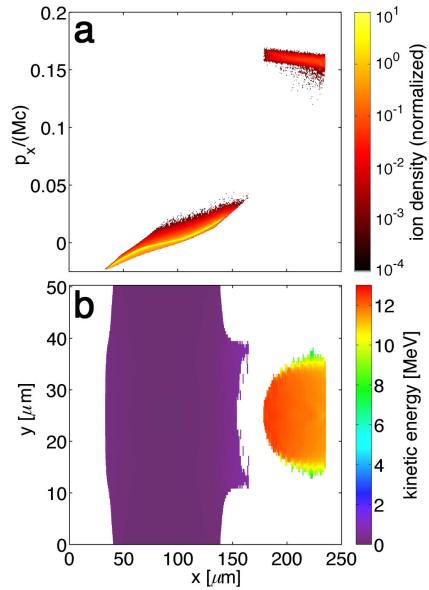, width=2.2in}
\caption{(Color online) (a) Ion distribution in the $x$-$p_x$ phase space and (b) mean kinetic energy of the ions as a function of position, after $6$ ps.}
\label{fig:phase_den}
\end{figure}

\begin{figure}[!htbp]
\centering \epsfig{file=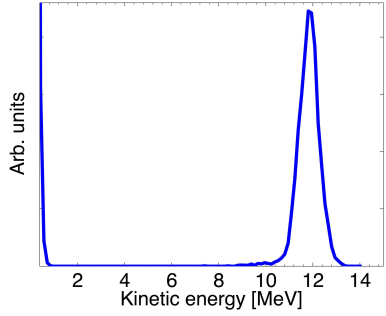, width=2.0in}
\caption{Distribution of proton kinetic energy after $6$ ps.}
\label{fig:spectrum}
\end{figure}

The trapping process begins after $1.8$ ps, with 4.2\% of the protons initially within the focal region of the lasers being extracted from the bulk distribution and accelerated by the ponderomotive beat wave. As the acceleration continues, the trapped protons form a bunch that separates from the main distribution both in space and in momentum, as illustrated in Fig. \ref{fig:phase_den}. After $6$ ps, the proton beam has a mean energy of $12$ MeV with a $7.5$\% energy spread (cf., Fig. \ref{fig:spectrum}, showing the proton spectrum), approximately $60$ $\mu$m long and $20$ $\mu$m wide, with a transverse momentum spread on the order of $10^{-2}$ $Mc$.
Much lower energy spreads, down to 1\% and below, can be obtained by employing longer pulses with the same energy and lower intensities (as discussed in Section \ref{sec:scalings}).
The expansion of the remaining plasma bulk, visible in Fig. \ref{fig:phase_den}a, is driven by the space charge that forms because of the rapid expansion of the hot electron population produced by the lasers (in the range of radiation intensity considered  here, the electron dynamics in the beat wave is dominated by stochastic heating \cite{heating_1,heating_2}).
These results confirm the potential of the proposed technique for direct acceleration as a method for producing high-quality ion beams in a controlled way.
Other simulation results, to be object of future work, suggest that, under appropriate conditions, higher plasma densities, generating significant space-charge fields, can even improve both trapping and acceleration.

\section{Conclusions}
\label{sec:conclusions}
The theory and the results presented here demonstrate the possibility of direct ion acceleration by the superimposed EM field of two counterpropagating, variable-frequency lasers with nonrelativistic radiation intensities. Calculations and simulations indicate that proof-of-principle experiments could already be performed using current ultraintense laser technology. 
The essential simplicity and robustness of the underlying physical mechanism, as well as its unique control capabilities, makes the present acceleration technique a promising candidate for the future development of laser-based ion accelerators, particularly when fine control over the beam features is required.

\section*{acknowledgments}
Work partially supported by FCT (Portugal) through grants POCI/FIS/55095/2003 and SFRH/BD/22059/2005, and by the European Community - New and Emerging Science and Technology Activity under the FP6 ``Structuring the European Research Area" programme (project EuroLEAP, contract number 028514).
The authors would like to acknowledge Prof. Tom Katsouleas and Dr. Igor Pogorelsky for stimulating discussions, and Michael Marti for help with the OSIRIS simulations, performed at the expp and IST clusters in Lisbon. 

\appendix[Validity of the ponderomotive approximation]
The squared vector potential, ${\bf A}^2 = ({\bf A}_1+{\bf A}_2)^2$, is given by
\begin{align}
  {\bf A}^2 = \ & \frac{A_1^2+A_2^2}{2} +  A_1A_2 \cos\left(\Phi\ped{S}\right) + \frac{\sin^2(\theta)-\cos^2(\theta)}{2} \Big[ 2A_1A_2 \cos\left(\Phi\ped{F}\right) + \nonumber\\
  &\left(A_1^2+A_2^2\right)\cos\left(\Phi\ped{S}\right)\cos\left(\Phi\ped{F}\right)-\left(A_1^2-A_2^2\right)\sin\left(\Phi\ped{S}\right)\sin\left(\Phi\ped{F}\right)\Big]\text{,}
\label{eq:Asquared}
\end{align}
where $\Phi\ped{F}(x,t)=\Phi_1+\Phi_2$ and $\Phi\ped{S}(x,t)=\Phi_1-\Phi_2$ indicate the phase of the fast and slow beat waves, respectively.
In order to obtain the ponderomotive equation \eqref{eq:x_motion_av}, the time average of the fast terms, containing $\Phi\ped{F}$, in Eq. \eqref{eq:Asquared} must be negligible, such that one can assume ${\bf A}^2=\frac{1}{2}\left(A_1^2+A_2^2\right)+A_1A_2\cos(\Phi\ped{S})$ [which is rigorously valid only for circular polarization, i.e., when $\sin^2\left(\theta\right)=\cos^2\left(\theta\right)$]. This occurs if the two following conditions are met: (i) the variation of $x$ and $p_x$ associated with the transverse quiver motion is negligible, which is verified if $\hat{A}_j = \frac{qA_j}{Mc^2}\ll1$ (here, $\hat{A}_j=\frac{Z\ped{i}m}{M}a_j$, where $a=\frac{eA}{mc^2}$ denotes the usual normalized vector potential, $e$ and $m$  being the electron charge and mass, and $Z\ped{i}$ is the ion charge state); (ii) along the ion trajectory $x\ped{i}(t)$, the frequency of the slow beat wave is much lower than the frequency of the fast beat wave, i.e., $\omega\ped{i}=-\frac{1}{2}\left[\frac{\partial\Phi\ped{S}}{\partial t}(x\ped{i},t)+v\ped{i}\frac{\partial\Phi\ped{S}}{\partial x}(x\ped{i},t)\right]$ is much lower than $\Omega\ped{i}=-\frac{1}{2}\left[\frac{\partial\Phi\ped{F}}{\partial t}(x\ped{i},t)+v\ped{i}\frac{\partial\Phi\ped{F}}{\partial x}(x\ped{i},t)\right]$, with $v\ped{i}(t)$ being the ion velocity. The latter condition is automatically satisfied for trapped ions, because their velocity stays always close to the phase velocity of the slow beat wave, namely, $v\ped{i}(t) \approx v_{\phi}[x\ped{i}(t),t]$, leading to
\begin{equation}
  \frac{\omega\ped{i}}{\Omega\ped{i}}\approx\frac{c(\Phi_1^{\prime}-\Phi_2^{\prime})-v_\phi(\Phi_1^{\prime}+\Phi_2^{\prime})}{c(\Phi_1^{\prime}+\Phi_2^{\prime})-v_\phi(\Phi_1^{\prime}-\Phi_2^{\prime})}\text{.}
\label{eq:dom_dOM}
\end{equation}
By recalling that $v_{\phi}=c\frac{\Phi_1^{\prime}-\Phi_2^{\prime}}{\Phi_1^{\prime}+\Phi_2^{\prime}}$,
one concludes that $\frac{\omega\ped{i}}{\Omega\ped{i}} \approx 0$ for any $t$; therefore, Eq. \eqref{eq:x_motion_av} provides an accurate description of the ion-trapping process, as well as of the long-term dynamics of trapped ions.

\end{document}